\documentclass[draft,grl]{agutex}
\usepackage{color}
\usepackage{graphicx}
\setkeys{Gin}{draft=false}
\authorrunninghead{SHUKLA ET AL.}
\titlerunninghead{Thermoelasticity of Iron-bearing bridgmanite}

\authoraddr{Corresponding author: Gaurav Shukla,
School of Physics and Astronomy, University of Minnesota, Minneapolis, Minnesota, USA.
(shukla@physics.umn.edu)}

\begin{document}

\title{Thermoelasticity of Fe$^{2+}$-bearing bridgmanite}

\authors{Gaurav Shukla,\altaffilmark{1} Zhongqing Wu,\altaffilmark{2} Han Hsu,\altaffilmark{3} Andrea Floris,\altaffilmark{4} Matteo Cococcioni,\altaffilmark{5} Renata M. Wentzcovitch,\altaffilmark{1,6}}

\altaffiltext{1}{School of Physics and Astronomy, University of Minnesota, Minneapolis, Minnesota, USA}

\altaffiltext{2}{School of Earth and Space Sciences, University of Science and Technology of China, Hefei, Anhui, China}

\altaffiltext{3}{Department of Physics, National Central University, Jhongli District, Taoyuan 32001, Taiwan}

\altaffiltext{4}{Department of Physics, King's College London, England, United Kingdom}

\altaffiltext{5}{Theory and Simulation of Materials (THEOS), \'{E}cole polytechnique f\'{e}d\'{e}rale de Lausanne, Station 12, CH-1015 Lausanne, Switzerland}

\altaffiltext{6}{Department of Chemical Engineering and Materials Science, University of Minnesota, Minneapolis, Minnesota, USA}

\date{\today}

\begin{abstract}
We present LDA+U calculations of high temperature elastic properties of bridgmanite with composition (Mg$_{(1-x)}$Fe$_{x}^{2+}$)SiO$_3$ for $0\le{x}\le0.125$. 
Results of elastic moduli and acoustic velocities for the Mg-end member (x=0) agree very well with the latest high pressure and high temperature experimental measurements. 
In the iron-bearing system, we focus particularly on the change in thermoelastic parameters across the state change that occurs in ferrous iron above $\sim$30 GPa, 
often attributed to a high-spin (HS) to intermediate spin (IS) crossover but explained by first principles calculations as a lateral displacement of substitutional
iron in the perovskite cage. We show that the measured effect of this change on the equation of state of this system can be explained
by the lateral displacement of substitutional iron, not by the HS to IS crossover. The calculated elastic properties of (Mg$_{0.875}$Fe$_{0.125}^{2+}$)SiO$_3$ 
along an adiabatic mantle geotherm, somewhat overestimate longitudinal velocities but produce densities and shear velocities quite consistent with 
Preliminary Reference Earth Model data throughout most of the lower mantle.
\end{abstract}

\begin{article}

\section{Introduction} \label{sec:intro}

Thermoelastic properties of lower mantle minerals provide a direct link to seismological observations.
In order to constrain the composition and thermal structure of the Earth's lower mantle, a detailed investigation 
of elastic properties of the constituent minerals is needed. Bridgmanite, (Mg,Fe,Al)(Si,Fe,Al)O$_3$ perovskite (Pv), is the main constituent of
the lower mantle along with ferropericlase, (Mg,Fe)O, CaSiO$_3$ perovskite, and (Mg,Fe,Al)(Si,Fe,Al)O$_3$ post-perovskite (PPv). Although there has 
been considerable progress in measurements of elastic properties of these minerals at high pressures and temperatures
\citep{Yeganeh-Haeri,Sinnelnikov,Fiquet,Andrault,Aizawa04,Jackson04,Jackson05b,Sinogeikin,Li1,Murakami1,Lundin,Ballaran,Chantel,Murakami2, Dorfman},
owing to extreme pressure and temperature conditions in the lower mantle the availability of data on Fe-bearing bridgmanite is quite limited to well
constrain lower mantle composition  and temperature. Several experiments and first-principles calculations have shown that pressure induced 
iron spin crossover \citep{Badro03,Badro04, Tsuchiya06, Lin12} affects elastic properties of (MgFe)O ferropericlase (fp) \citep{Goncharov,Crowhurst,Marquardt09a,Wentzcovitch09,Wu09,Antonangeli11,Wu13,Wu14}.
In the case of iron-bearing bridgmanite, effects of spin crossover on elastic properties have been quite uncertain due to possible 
coexistence of ferrous (Fe$^{2+}$) and ferric (Fe$^{3+}$) iron along with the more complex perovskite crystal structure.
Experimentally, it has been difficult to isolate the effects of ferrous and ferric iron unambiguously.  

Acoustic velocity measurements using an ultrasonic technique \citep{Chantel} and compression curve measurements \citep{Ballaran} in
iron-bearing bridgmanite, demonstrated a significant change in bulk modulus across the state change in iron. By fitting experimental data 
with 3$^{rd}$-order Birch-Murnaghan equation
of state to small (0-40 GPa) and large (0-75 GPa) pressure ranges, \citet{Ballaran} found ambient condition bulk modulus smaller (245 GPa) 
and larger (253 GPa), respectively. Since these fitting pressure ranges had no effect on the iron-free phase, \citet{Ballaran} attributed this behavior
to change in compression mechanism caused by the high-spin (HS) to intermediate-spin (IS) crossover \citep{McCammon}.
However the HS to IS crossover has not been found in first-principle studies \citep{Bengtson09,Hsu10,Hsu11,Hsu14}. 
Using the local density approximation augmented by the Hubbard type correction (LDA+U) method,  
\citet{Hsu10} showed that iron in Fe$^{2+}$-bearing bridgmanite should remain in the HS state throughout the lower mantle 
and should, instead, undergo a pressure induced displacement 
producing a state with increased iron M$\ddot{o}$ssbauer quadrupole splitting (QS).
The calculated low (1.9-2.4 mm/s) and high (3.3-3.9 mm/s) QS states were consistent with experimental measurements \citep{Lin12,McCammon,McCammon13,
potapkin13, kupenko14}. 
Owing to lack of sufficient experimental measurements on elasticity of iron-bearing bridgmanite,
effects of the proposed HS to IS crossover \citep{McCammon}, or iron displacement \citep{Bengtson09,Hsu10,Hsu11, Hsu14}, have not been properly understood.
To clarify this issue we have calculated the high temperature and high-pressure elastic moduli and acoustic velocities of (Mg$_{1-x}$Fe$_x$)SiO$_3$ for x=0 and x=0.125,
with iron in low- and high-QS states. We also compare the effects of the HS to IS transition \citep{Hsu14} in the static equation of state of Fe$^{2+}$-bearing bridgmanite. 

Unlike in (MgFe$^{3+}$)(SiFe$^{3+}$)O$_3$ \citep{Hsu11}, the high-spin (HS) to low-spin (LS) transition in (MgFe$^{2+}$)SiO$_3$ has not been found in the lower-mantle pressure range using the LDA+U method \citep{Hsu10,Hsu14}. 
However, there have been several studies suggesting that HS to LS transition might occur at around $\sim$ 90 GPa \citep{Badro04,Li04,Li06,Jackson05a,McCammon,Umemoto08,Lin12} and
might have consequences for the vibrational spectrum \citep{Caracas14}. Recently \citet{Zhang14} found that iron-bearing bridgmanite dissociates into nearly iron-free phase and into an iron-rich hexagonal 
silicate phase in the deep lower mantle. As the dissociation phase boundary is still not well known, we present results for elasticity of Fe$^{2+}$-bearing bridgmanite with constant iron concentration 
in the entire pressure range of the lower mantle. 

\section{Method and Calculation Details} \label{sec:method}

We have used density functional theory (DFT) within the local density approximation (LDA) \citep{ceperley} augmented by the Hubbard U (LDA+U) calculated 
self-consistently \citep{Cococcioni,Kulik, Hsu09, Leiria-Campo10, Himmetoglu14}.
Ultrasoft pseudo-potentials \citep{Vanderbilt} have been used for
Fe, Si, and O. For Mg, a norm-conserving pseudo-potential generated by von Barth-Car's method has been used.  A detailed description about pseudo-potentials is available in \citet{Umemoto08}.
The plane-wave kinetic energy and charge density cut-off are 40 Ry and 160 Ry, respectively. 
Structural optimization at arbitrary volumes has been performed using variable cell-shape damped molecular dynamics
\citep{Wentzcovitch91,Wentzcovitch93}. Self- and structurally-consistent U$_{sc}$=3.1 eV, reported by \citet{Hsu10} using the linear response approach \citep{Cococcioni,Kulik}, 
has been used for high-spin low-QS and high-QS iron.
Vibrational density of states (VDoS), needed to calculate free energy 
within quasi-harmonic approximation (QHA) \citep{Carrier07}, have been calculated using density functional perturbation theory (DFPT) \citep{Baroni} using the LDA+U functional \citep{Floris}. 
(Mg$_{1-x}$Fe$_x$)SiO$_3$ for x=0 and 0.125 have been 
investigated in a 40-atom super-cell.  Elastic properties for  0\textless{x}\textless{0.125} are linearly interpolated using x=0 and x=0.125 results.  Electronic states have been sampled on $2\times2\times2$ 
Monkhorst-Pack k-point grid  shifted by $(1/2,1/2,1/2)$ from the Brillouin Zone origin. 
VDoS have been obtained by calculating dynamical matrices on a $2\times2\times2$ q-point grid and interpolating the obtained force constants on a $8\times8\times8$ q-point grid.
The structure for x=0, and x=0.125 with high-QS state have been optimized at twelve pressures between -10 and 150 GPa.
The low-QS structure has been optimized up to 60 GPa only. Beyond this pressure unstable phonons appear.
To obtain static elastic coefficients (zero temperature), C$_{ij}$, at each pressure, positive and negative strains of 1$\%$ magnitude have been applied and after relaxing the internal
degrees of freedom, associated stresses have been calculated. Thermoelastic moduli have been calculated using a semi-analytical approach \citep{Wu11}. 
This method uses an analytical expression for strain Gr$\ddot{u}$neisen parameters to calculate the thermal contribution to elastic coefficients using the quasi-harmonic approximation. 
This method allows calculations of thermal elastic coefficients using static values and vibrational density of states for unstrained configurations only.
It is almost two orders of magnitude more efficient than the fully numerical method \citep{karki,Wentzcovitch04}, which required vibrational density of states also for strained configurations.
It also appears to be more accurate since it avoids numerical differentiation on a reduced number of grid points as well as calculations of VDoS and free energies for strained configurations.
The method has been applied successfully to several minerals already \citep{Wu11,Nunez-Valdez12a,Nunez-Valdez12b,Nunez-Valdez13,Yang14}. Calculations have been performed using VLab cyberinfrastructure at the Minnesota Supercomputing Institute \citep{pedro08}.

\section{Results and Discussion} \label{sec:result}

\subsection{Elasticity of Iron-free and Fe$^{2+}$-bearing Bridgmanite} \label {elas}

\begin{figure}
\includegraphics[width=15cm]{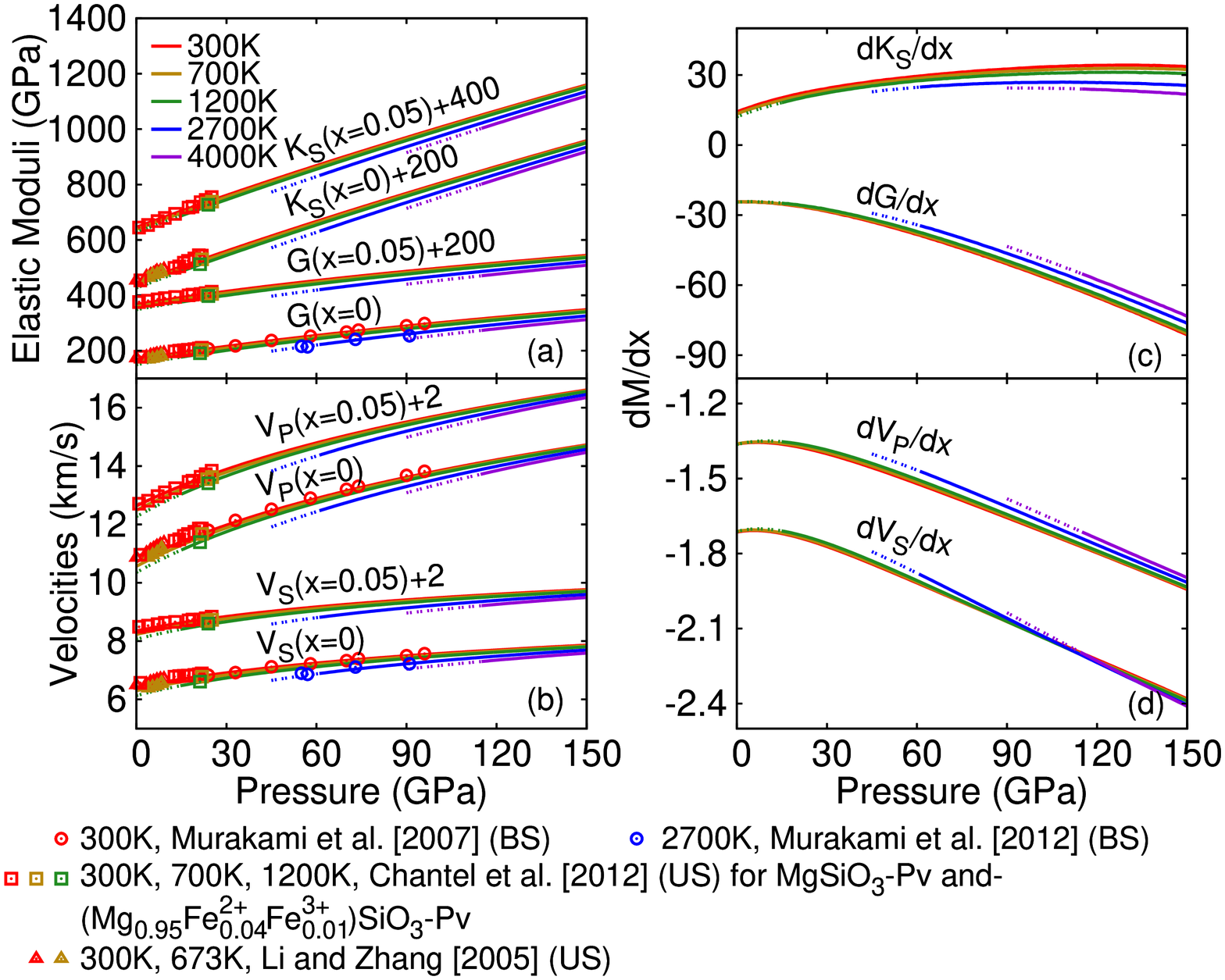}
\caption{(Color online) Pressure and temperature dependence of elastic moduli and sound velocities for
 (Mg$_{1-x}$Fe$_x^{2+}$)SiO$_3$ with x=0 and 0.05 (a, b), and of dM/dx (M= K$_S$, G, V$_P$, V$_S$) 
 (c, d). Solid (dashed) lines represent first-principles results within (outside)
the validity of quasi-harmonic approximation. BS: Brillouin scattering, US: Ultrasonic technique.}
\label{fig.elas.Pv}
\end{figure}

\begin{figure}
\includegraphics[width=12cm]{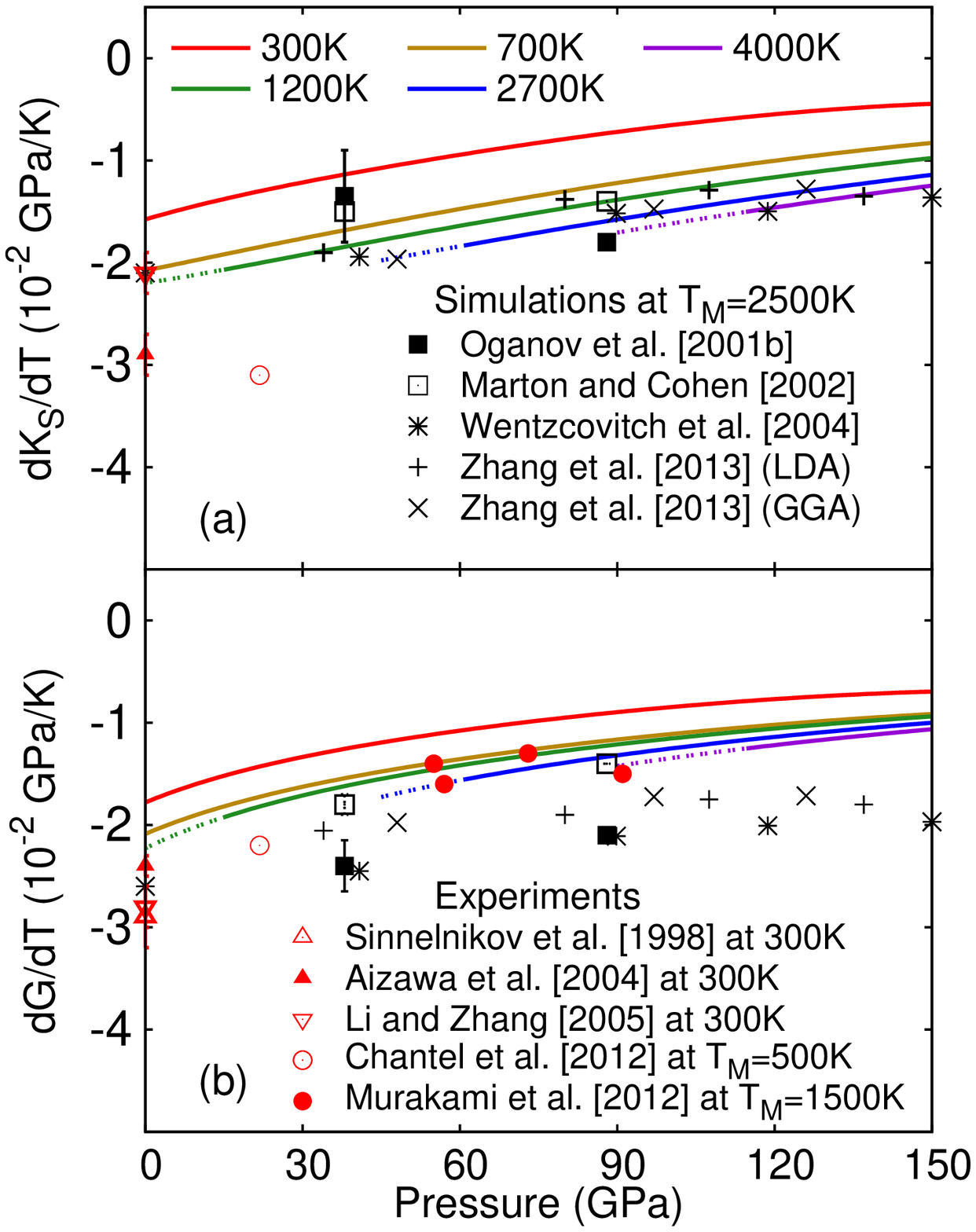}
\caption{(Color online) Pressure and temperature dependence of (a) $\frac{dK_S}{dT}$, and (b) $\frac{dG}{dT}$ for iron-free bridgmanite. 
Results from this study (lines) are compared with previous simulations (black symbols) and experimental data (red symbols).
The values of $\frac{dK_S}{dT}$ and $\frac{dG}{dT}$ obtained from previous simulations \citep{Oganov01b,Marton,Wentzcovitch04,Zhang13} and experimental
measurements \citep{Chantel, Murakami2} shown here are calculated at the midpoint between extreme temperatures
[i.e., $\frac{dM(P,T_M)}{dT}$=$\frac{M(P,T_2)-M(P,T_1)}{T_2-T_1}$, 
where $T_M=\frac{T_1+T_2}{2}$ and M=(K$_S$, G)]. Error bars on \citet{Oganov01b} and \citet{Marton} results are reproduced from \citet{Zhang13} while on \citet{Wentzcovitch04} and \citet{Zhang13}
they are smaller than the symbols size.}
\label{fig.Temp-deriv.Pv}
\end{figure}

The calculated aggregate elastic moduli, K and G, and acoustic velocities, V$_P$ and V$_S$, of iron-free bridgmanite compare  well with Brillouin scattering measurements of 
\citet{Murakami1,Murakami2} as shown in Fig. \ref{fig.elas.Pv}(a) and \ref{fig.elas.Pv}(b). Within experimental uncertainties,
ultrasonic measurements of \citet{Li1} and \citet{Chantel} also compare well with our results. Table \ref{table:elastic-data} 
compares calculated elastic properties at zero pressure and 300 K with 
measurements and simulations  at ambient conditions. As compared to experimental measurements, our calculated volume is overestimated by $\sim$ 0.5$\%$
while elastic moduli and acoustic velocities are underestimated by $\sim$ 1-3$\%$ , which is typical for LDA calculations after including zero-point energy \citep{karki,Karki01, Wentzcovitch04, Wentzcovitch10a, Wentzcovitch10b}.
Calculated pressure derivatives of elastic moduli, K$'$ and G$'$, are 3.96 and 1.79, respectively. Our K$'$ agrees reasonably well with previous studies while G$'$ compares well with those reported by  
\citet{Sinnelnikov} and \citet{Zhang13}. 
Comparing acoustic velocities
and elastic moduli at higher temperatures, especially shear velocities and shear moduli at 2,700 K by \citet{Murakami2}, 
we find that the temperature dependence of elastic properties has been captured well. 
As shown in Fig. \ref{fig.Temp-deriv.Pv}, our calculated $\frac{dK_S}{dT}$  and $\frac{dG}{dT}$ are smaller in magnitude compared 
to values obtained by ultrasonic \citep{Sinnelnikov,Aizawa04,Li1} at $0$ GPa and  Brillouin scattering measurements \citep{Chantel} at $22$ GPa.  Calculated values of $\frac{dK_S}{dT}$  
compare well with those of previous calculations \citep{Oganov01b,Marton,Wentzcovitch04, Zhang13}.
Calculated $\frac{dG}{dT}$ agrees well with Brillouin scattering measurements by \citet{Murakami2} and non-self consistent calculation by \citet{Marton} 
but differs substantially from that of \citet{Oganov01b}, \citet{Wentzcovitch04}, and \citet{Zhang13}.

The semi-analytical method used here had produced well the temperature dependence of elasticity for all applied materials
\citep{Wu11,Nunez-Valdez12a,Nunez-Valdez12b,Nunez-Valdez13, Yang14}. This method has also produced slightly better results for MgO \citep{Wu11} than the fully numerical  method
\citep{karki,Wentzcovitch04}. This is because this method avoids calculations of vibrational frequencies in 
strained configurations and numerical errors associated with finite difference calculations of elastic coefficients. ‍

The elastic moduli and acoustic velocities of iron-bearing bridgmanite at lower mantle pressures are also shown in  Fig. \ref{fig.elas.Pv}.   
To compare with measurements of \citet{Ballaran} and \citet{Chantel} in the iron-bearing phase, we estimated elastic properties for 0\textless{x}\textless{0.125} by linearly interpolating our
calculated results for x=0 and x=0.125. The calculated volume at ambient conditions is in agreement with measurements by \citet{Ballaran} for x=0.04 and is smaller by  $\sim$2$\%$  when compared 
with measurements by \citet{Chantel} for x=0.05 (Table \ref{table:elastic-data}). 
As shown in Fig. \ref{fig.elas.Pv}(b), comparison of calculated acoustic velocities, V$_P$ and V$_S$, for x=0.05 with the only available experimental measurements for 
(Mg$_{0.95}$Fe$^{2+}_{0.04}$Fe$^{3+}_{0.01}$)SiO$_3$ \citep{Chantel} is also good except for the shear velocity data in the lower pressure range. 
As seen in Fig. \ref{fig.elas.Pv} (c) and (d) and in Table \ref{table:elastic-data}, the presence of iron has very little 
to no effect on bulk modulus, K$_S$, but shear modulus G  softens by $\sim$2 to 6$\%$, consistent with previous static first-principles studies \citep{Kiefer,Lundin}. 
The effect of iron on the temperature dependence of acoustic velocities and elastic moduli is not very significant (see Fig. \ref{fig.elas.Pv}).

\subsection{Effect of Pressure Induced Iron Displacement on Elasticity} \label {effect-low-QS}

\begin{figure}
\includegraphics[width=12cm]{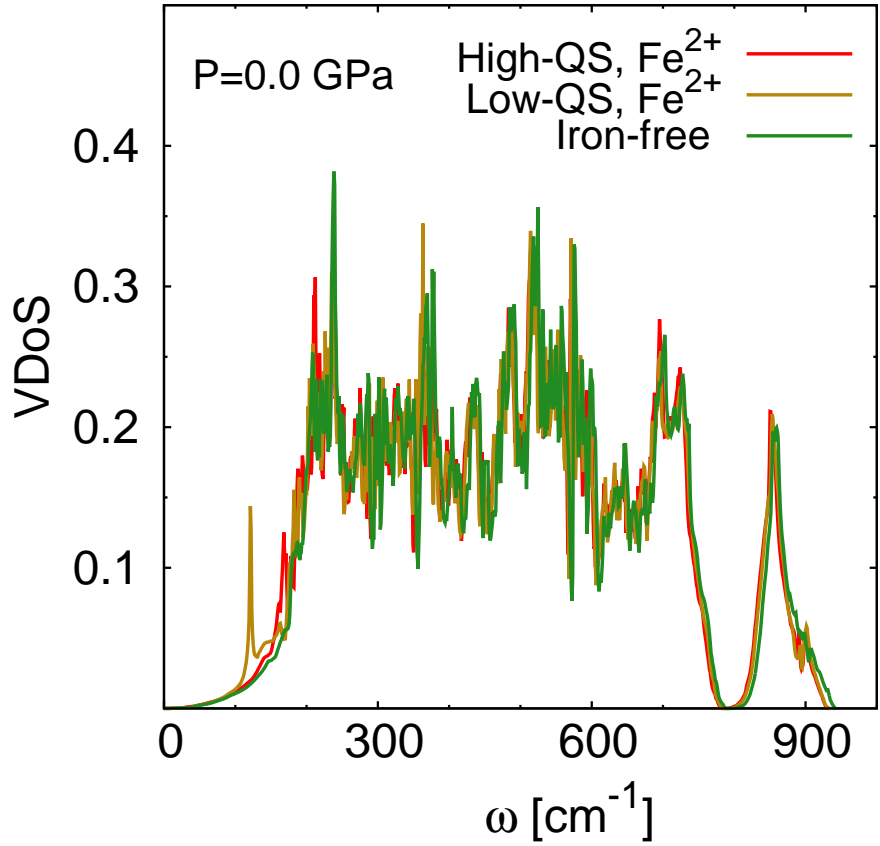}
\caption{(Color online) Vibrational density of states (VDoS) at $P=0$ for iron-free bridgmanite using LDA,
 and for low and high QS state of (Mg$_{0.875}$Fe$_{0.125}^{2+}$)SiO$_3$  using LDA+U.}
\label{fig.v-DOS}
\end{figure}

We have investigated the elastic consequences of the pressure induced iron displacement \citep{Hsu10,Hsu14} by calculating thermal elastic moduli and acoustic velocities 
for x=0.125 with iron in low-QS and high-QS sites.  The low-QS structure is stable up to 60 GPa in LDA+U calculations. Above this pressure unstable phonon modes start appearing.
As shown in Fig. \ref{fig.v-DOS}, vibrational density of states (VDoS) for x=0 and for x=0.125 with iron in the high-QS and low-QS sites are not very different except for a small shift towards
lower frequencies owing to the presence of iron. However, there is an extra peak at $\sim$ 150 cm$^{-1}$ in the
VDoS if iron is in the low-QS site. It is worth noting that the vibrational mode associated with this extra peak in the low-QS structure is infra-red active 
while the same mode is infra-red inactive in the high-QS structure. This information may provide a crucial test to investigate the state of ferrous iron in bridgmanite.

\begin{figure}
\includegraphics[width=8cm]{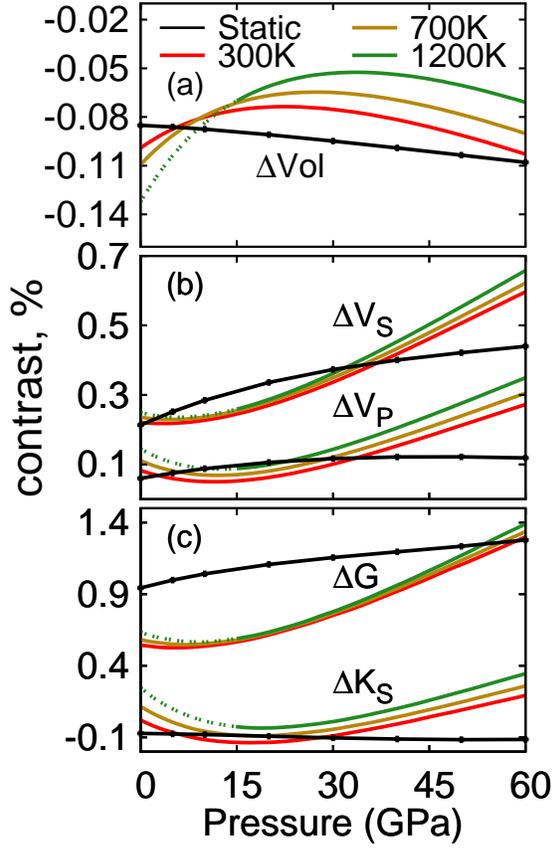}
\caption{(Color online) Contrast (${\Delta}M$)  in (a) unit-cell volume, (b) 
compressional and shear velocity ($V_P$, $V_S$), and (c) bulk and shear modulus ($K_S$, $G$) between low-QS and high-QS states in (Mg$_{0.875}$Fe$_{0.125}^{2+}$)SiO$_3$.}
\label{fig.Del-Vel}
\end{figure}

The effect of iron displacement (low-QS to high-QS transition) on unit-cell volume, acoustic velocities, and elastic moduli of iron-bearing bridgmanite with x=0.125 is shown in Fig. \ref{fig.Del-Vel}.
To quantify the magnitude of the property change, we define the contrast $\Delta$ in a particular property M as 
${\Delta}M=\frac{M_{high-QS}-M_{low-QS}}{M_{low-QS}}\times100$. As shown in Fig. \ref{fig.Del-Vel}, the volume contrast  is very small ($\sim$0.1$\%$), consistent
with \citet{Lundin} and \citet{Ballaran}, and is unlikely to be detected in experiments.
The contrast in bulk modulus, K$_S$, and acoustic velocities, V$_P$ and V$_S$, are also small. However, the shear modulus contrast is about $\sim$0.5-1.5$\%$. 
This change in G, was also suggested by \citet{Ballaran} using spontaneous strain analysis. 
To investigate the fitting pressure-range behavior observed by \citet{Ballaran} and \citet{Chantel} the calculated results for low-QS and high-QS  have been combined at 30 GPa, the pressure above which the HS to IS
crossover was proposed by  \citet{McCammon} and \citet{ potapkin13}. The 3$^{rd}$-order Birch-Murnaghan equation of state was used to fit these combined results. 
As shown in Table \ref{table:fitting-behavior}, depending on the pressure-range of the combined results, an isothermal bulk modulus softening of $\sim$3 to 7 GPa has been found. 
This softening, however, reduces with increasing temperature. 

Intermediate spin (IS) states of the iron in Fe$^{2+}$-bearing bridgmanite are energetically unfavored in the lower mantle pressure-range \citep{Hsu10,Hsu14} and phonon calculations have
been extremely difficult for these states. To understand the overall trend of fitting parameters of combined low-QS and IS data, we have used static data 
for low-QS state and IS state with QS value 0.9-1.6 mm/s \citep{Hsu14}. 
The static bulk modulus at zero pressure for low-QS and IS states are 263.2 GPa and 254.1 GPa, respectively (Figure S3). When static low-QS and IS results have been combined at 30 GPa 
and fitted within different pressure-ranges of 0-150 GPa and 0-120 GPa, static bulk modulus of the combined fit are 305 GPa and 348.7 GPa, respectively (Figure S4).
 The fitting pressure-range trend of combined low-QS and high-QS is the same as that observed by \citet{Ballaran} and \citet{Chantel} while of combined low-QS and 
IS states the fitting trend is opposite. These findings substantiate our understanding that ferrous 
iron in perovskite remains in the HS state in the lower mantle pressure range, but instead, undergoes a horizontal displacement in the perovskite cage which results in an increase in M$\ddot{o}$ssbauer quadrupole splitting. 

\section{Geophysical Significance} \label {Geophys}

\begin{figure}
\includegraphics[width=15cm]{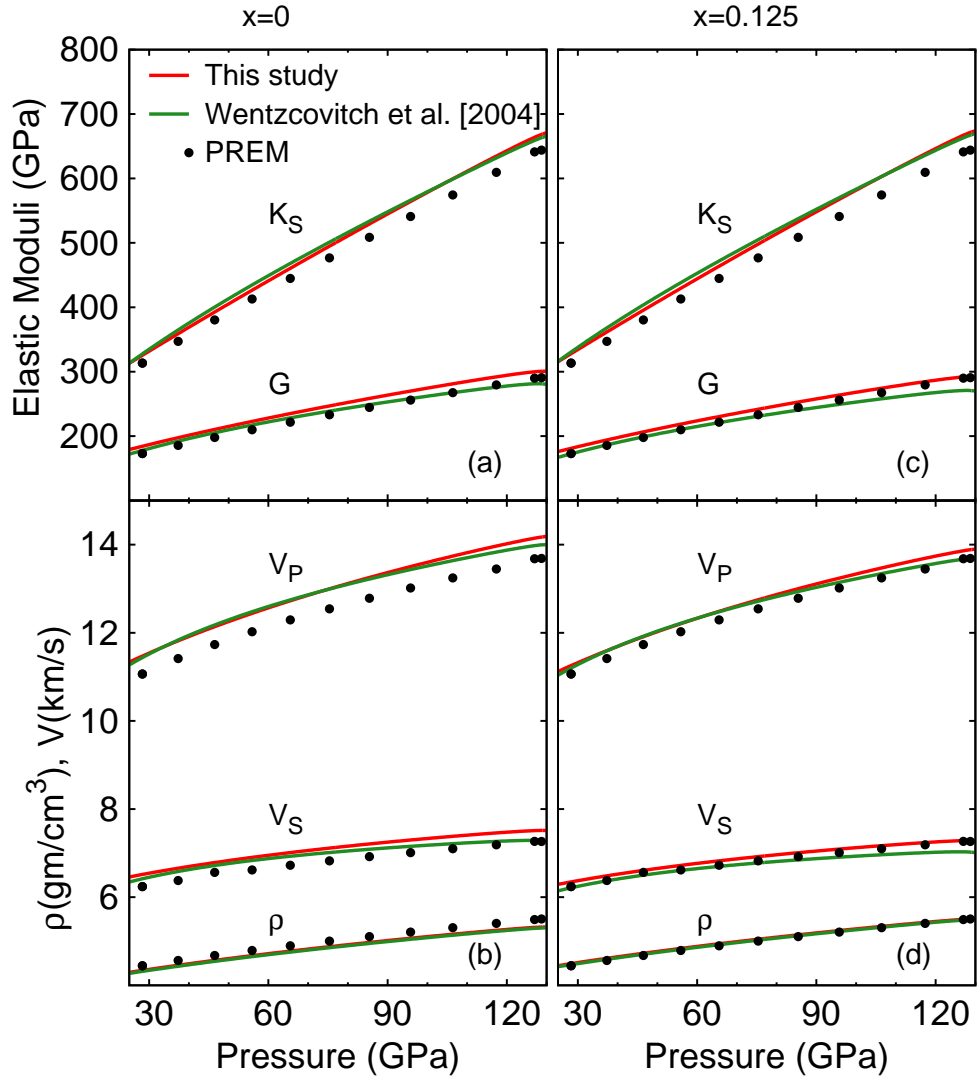}
\caption{(Color online) Elastic moduli ($K_S$, $G$), sound velocities ($V_P$, $V_S$), and density ($\rho$) 
 for (Mg$_{1-x}$Fe$_{x}^{2+}$)SiO$_3$ with x=0 (a, b) and x=0.125 (c, d) calculated along the Boehler's geotherm \citep{Boehler00} using present \citep{Wu11} and previous \citep{Wentzcovitch04} methods.}  
\label{fig.PREM}
\end{figure}

As shown in Fig. \ref{fig.PREM}, elastic moduli ($K_S$, $G$), acoustic velocities ($V_P$, $V_S$), and densities
($\rho$) are calculated along the 
Boehler's geotherm \citep{Boehler00} using new \citep{Wu11} and previous \citep{karki, Wentzcovitch04} 
methods and the results are compared with seismic values from the Preliminary Reference Earth Model (PREM) 
\citep{Dziewonski}. Owing to the smaller magnitude of $\frac{dG}{dT}$ obtained here, present results for G and Vs differ more from previous results by \citet{Wentzcovitch04} in the deep lower mantle region,
where temperatures are higher. This difference is sufficient to alter conclusions regarding mantle composition.    
Inclusion of 12.5$\%$ Fe$^{2+}$ in MgSiO$_3$ perovskite produces good agreement of $\rho$, $V_S$, and $G$ with PREM values (Fig. \ref{fig.PREM}c,d).
Therefore, comparison of calculated $V_S$ with PREM's $V_S$ may indeed suggest a more perovskitic lower manle \citep{Murakami2}. However,
relatively large values of $K_S$ and $V_P$ are found, suggesting that the lower mantle may accommodate a reasonable amount of ferropericlase, (MgFe)O, and CaSiO$_3$ pereovskite as previously 
indicated \citep{Wentzcovitch04}. Resolving whether the lower mantle composition is pyrolytic or perovskitic still needs more detailed investigations 
of the effect of Al$_2$O$_3$ and Fe$_2$O$_3$ on the elasticity of bridgmanite along with the elasticity of CaSiO$_3$ perovskite. 
Ultimately, the analysis of one dimensional velocity profiles might be too limited to address this question more conclusively. 
Analysis of lateral heterogeneities, including effects of spin crossovers in (MgFe)O \citep{Wu14} and in bridgmanite might prove more useful to address this question.

\section{Conclusions} \label{sec:conc}

We have presented first-principles LDA+U calculations of aggregate elastic moduli and acoustic velocities for Fe$^{2+}$-bearing bridgmanite at lower mantle conditions.
Using the new semi-analytical approach for high temperature elasticity \citep{Wu11}, we find an unexpected difference of temperature gradients 
with respect to previous fully numerical calculations \citep{Wentzcovitch04} also based on the quasi-harmonic approximation. New results for $\frac{dG}{dT}$ and $\frac{dV_S}{dT}$ 
for iron-free bridgmanite are in good agreement with the latest high temperature high pressure measurements of these quantities by \citet{Murakami2}, which also differ from previous calculations and previous experiments.
Overall, it appears that pressure and temperature gradients of bulk and shear moduli are better estimated now.
Calculated acoustic velocities for iron-bearing bridgmanite also agree well with ultrasonic measurements of \citet{Chantel}. 
We do not find any significant effect of iron-site change (low-QS to high-QS) on sound velocities and elastic moduli, except for $\sim$1$\%$ increase in shear modulus, consistent with findings by \citet{Ballaran}. 
Effects of fitting combined data for low-QS and high-QS states on elastic properties are consistent with previously observed fitting pressure-range behavior observed by \citet{Ballaran} and \citet{Chantel}. 
Calculated shear modulus and shear velocities of bridgmanite along a typical mantle geotherm are qualitatively more consistent with PREM values now 
compared with previous first principles results \citep{Wentzcovitch04}. Differences in bulk modulus and longitudinal velocities remain noticeable. 

\begin{acknowledgement}
This work was supported primarily by grants NSF/EAR 1319368 and NSF/CAREER 1151738. Han Hsu was supported by NSC 102-2112-M-008-001-MY3.
Computations were performed at the Minnesota Supercomputing Institute (MSI) and at the Blue Waters System at NCSA. 
The data for this paper are available at http://www.vlab.msi.umn.edu/reports/gauravpublications/index.shtml
\end{acknowledgement}

\end{article}

\newpage

\begin{table}
\caption{Volume, sound velocities, elastic moduli and their pressure derivatives at ambient pressure 
and temperature conditions. BS: Brillouin scattering, US: Ultrasonic, XRD: X-ray diffraction. }
\centering
\label{table:elastic-data}
\resizebox{18cm}{!} {
\begin{tabular}{ l l l l l l l l p{7.8cm}}
\hline
$V(\mathring{A}^3$) & x &$V_P(km/s)$ & $V_S(km/s)$ & $K_S$(GPa) & $G$(GPa) & $K'$ & $G'$ & References \\
\hline
\hline
163.2 & 0 & 10.72  &  6.42 & 245.5 & 168.2 & 3.96 & 1.79 & This study, single crystal/LDA \\
163.3 & 0.05 & 10.63 & 6.34 & 245.8 & 167.02 & 4.03 & 1.79 & This study, single crystal/LDA+U (Low-QS)\\
162.3(0.6) & 0 & 11.04 & 6.57 & 264(5) & 177(4) & - & - & \citet{Yeganeh-Haeri}, sinlge-crystal/BS\\ 
162.3 & 0 & - & 6.51 & - & 176(5) & -& 1.8(0.4)& \citet{Sinnelnikov}, poly-crystal, US\\
162.4(0.5) & 0 & 10.84(0.1)& 6.47(0.06) & 253(5) & 173(3) & - & - & \citet{Sinogeikin}, poly-crystal/BS\\
162.4(0.5) & 0 & 10.88(0.06) & 6.54(0.03) & 253(3) & 175(2) & - & -& \citet{Sinogeikin}, sinlge-crystal/BS\\
162.3 & 0 & 10.86(0.05) & 6.49(0.04) & 253(2) & 173(1) & 4.4(0.1) & 2.0(0.1) & \citet{Li1}, poly-crystal/US\\
- & 0 & 10.85(0.03) & 6.49(0.03) & - & 172.9(1.5) & - & 1.56(0.04) & \citet{Murakami1}, poly-crystal/BS\\
162.2(0.2) & 0 & 10.82(0.06) & 6.54(0.04) & 247(4) & 176(2) & 4.5(0.2) & 1.6(0.1) & \citet{Chantel}, poly-crystal/US\\
166.5(0.2) & 0.05 & 10.60(0.06) & 6.46(0.04) &  236(2) & 174(1) & 4.7(0.1) & 1.56(0.5) & \citet{Chantel}, poly-crystal/US\\
162.4 & 0 & - & - & 267 & 180 & 4.10 & - & \citet{Oganov01b}, Simulations\\
162.49 & 0 & - & - & 269(2) & 159(2) & - & - & \citet{Marton}, Simulations \\
162.4 & 0 & - & - & 250.5 & 172.9 & 4.01 & 1.74 & \citet{Zhang13}, Simulations\\
\hline
Compression \\
Studies\\ 
$V(\mathring{A}^3$) & & & & K$_T$ & & $K'$ &  &\\
162.3 & 0 & & & 259.5 & & 3.69 & & \citet{Fiquet}\\
162.4 & 0 & & & 248.8 & &  4 & & \citet{Andrault}\\
162.7 & 0.05 & & & 255.4 & &  4 & & \citet{Andrault}\\
164.1 & 0 & & & 247 & & 3.97 & & \citet{Karki01}, Simulations \\
162.30 & 0 & & & 255-261 & & 4 & & \cite{Lundin}\\
162.18 & 0.09 & & & 257-259 & & 4 & & \cite{Lundin}\\
163.30 & 0.15 & & & 257-259 & & 4 & & \cite{Lundin}\\
163.09(0.06) & 0.04 & &  & 253(2) &  & 3.99(0.07) &  & \citet{Ballaran},  (0-75 GPa)\\
163.16(0.06) & 0.04 & &  & 245(4) &  & 4.4(0.03) &  & \citet{Ballaran}, (0-40 GPa)\\
166.5(0.2) & 0.05 &  &  & 246(2) & & 4 & & \citet{Chantel}, \\
163.0 & 0.09 & &  & 251(13) &  & 4 &  & \citet{Dorfman} \\
\hline
\hline
\end{tabular}
}
\end{table}

\begin{table}
\caption{Calculated equation of state parameters for Low-QS, High-QS, and combined results at 30 GPa for low-QS and high-QS fitted with different pressure range.} \centering
\label{table:fitting-behavior}
\begin{tabular}{ p{5.0cm} | c c c | c c c}
\hline
&  \multicolumn{3}{|c|}{300K} &  \multicolumn{3}{|c}{700K} \\
\hline
& $V(\mathring{A}^3$)  & K$_T$ & K$'$ & $V(\mathring{A}^3$)  & K$_T$ & K$'$ \\
Low-QS & 163.81  & 247.16  & 4.05 & 165.62  & 234.05  & 4.12\\
High-QS & 163.64  & 248.90  & 3.97 & 165.44  & 236.44  & 4.03\\
Combined data (0-150 GPa) & 163.81  & 245.59  & 4.02 & 165.62  & 233.93  & 4.06 \\
Combined data (0-90 GPa) & 163.87  & 243.41  & 4.09 & 165.67  & 232.32  & 4.11 \\
Combined data (0-75 GPa) & 163.93 & 241.25 & 4.15 & 165.68 & 232.29 & 4.12 \\
\hline
\end{tabular}
\end{table}

\end{document}